\documentclass{aa} 
\usepackage{graphicx}
\usepackage{txfonts}
\usepackage[authoryear]{natbib}
\usepackage{lscape}                                
\usepackage{color}
\newcommand{\hii} {H\,{\sc ii}}
\newcommand{\Teff} {$T_{\rm eff}$}
\newcommand{\grav} {log\,{\em g}}
\newcommand{\vsini} {$v$\,sin\,$i$}
\newcommand{\micro} {$\zeta_{\rm t}$}
\newcommand{\kms} {km\,s$^{-1}$}

\newcommand{\fastwind} {FASTWIND}

\newcommand{\cloudy} {Cloudy}
\newcommand{\idl} {{\sc IDL}}

\newcommand{\solu}[2]{#1\,$\pm$\,#2}

\newcommand{\iraf} {{\sc IRAF}}
\newcommand{\fwhm} {{\sc fwhm}}
\newcommand{\newton}{INT}
\newcommand{\herschel}{WHT}
\newcommand{\ids}{IDS}
\newcommand{\isis}{ISIS}
\newcommand{\wfc}{WFC}

\newcommand{\ha}{{H$\alpha$}}
\newcommand{\hb}{{H$\beta$}}

\newcommand{\Te}{$T_{\rm e}$}
\newcommand{\Ne}{$n_{\rm e}$}
\newcommand{\te}{$T_{\rm e}$}

\newcommand{\foiii}{[O~{\sc iii}]}

\newcommand{\foi}{[O~{\sc i}]}
\newcommand{\foii}{[O~{\sc ii}]}
\newcommand{\fsii}{[S~{\sc ii}]}
\newcommand{\fsiii}{[S~{\sc iii}]}

\newcommand{\fnii}{[N~{\sc ii}]}
\newcommand{\fneiii}{[Ne~{\sc iii}]}
\newcommand{\fariii}{[Ar~{\sc iii}]}
\newcommand{\hi}{H\,{\sc i}}
\newcommand{\hei}{He~{\sc i}}

\begin{document}
%
\title{A detailed study of the \ion{H}{ii} region M\,43 and its
ionizing star\thanks{The \newton\ and \herschel\ telescopes are operated 
		  on the island of La Palma by the RGO in the Spanish 
		   Roque de los Muchachos Observatory of the Instituto 
		  de Astrof\'{\i}sica de Canarias.}}

\subtitle{I. Stellar parameters and nebular empirical analysis}

\author{S. Sim\'on-D\'{\i}az\inst{1,2},  J. Garc\'ia-Rojas\inst{1,2,3}, C. Esteban\inst{1,2},  G. Stasi\'nska\inst{4},
A.\,R. L\'opez-S\'anchez\inst{5,6} \& C. Morisset\inst{3}}

\institute{Instituto de Astrof\'\i sica de Canarias, E-38200 
           La Laguna, Tenerife, Spain
	   \and
           Departamento de Astrof\'isica, Universidad de La Laguna, E-38205 
           La Laguna, Tenerife, Spain.
	   \and
           Instituto de Astronom\'ia, Universidad Nacional Aut\'onoma de M\'exico, 
           M\'exico, D.F., M\'exico.
           \and
	   LUTH, Observatoire de Paris, CNRS, Universit\'e Paris Diderot; 
           5 Place Jules Janssen, F-92190 Meudon, France
           \and
           Australian Astronomical Observatory, P.O. Box 296, Epping, NSW 1710, Australia
           \and
           Department of Physics and Astronomy, Macquarie University, NSW 2109, Australia 
	   }
\offprints{ssimon@iac.es}

\date{Submitted/Accepted}
\titlerunning{M43}
\authorrunning{S. Sim\'on-D\'{\i}az et al.}

%
\abstract
{}
{We have selected the Galactic \hii\ region M\,43, a close-by apparently spherical
nebula ionized by a single star (HD37061, B0.5\,V) to investigate 
several topics of recent interest in the field of \hii\ regions and massive 
stars. In a series of two papers, we perform a combined, comprehensive 
study of the nebula and its ionizing star by using as many observational constraints
as possible. 
}
{We collected for this study a set of high-quality observations, including the 
optical spectrum of HD\,37061, along with nebular optical 
imaging and long-slit spatially resolved spectroscopy. The first part of
our study comprises a quantitative spectroscopic analysis of the ionizing star, 
and the empirical analysis of the nebular images and spectroscopy. All the 
information obtained here will be used to construct a customised photoionization 
model of the nebula in Paper II.}
{We determine the stellar parameters of HD\,37061 and the total number of ionizing 
photons emitted by the star. We find observational evidence  
of scattered light from the Huygens region (the brightest part of the Orion nebula)
in the M\,43 region. We show the importance of an adequate correction
of this scattered light in both the imagery and spectroscopic observations of M\,43
in accurately determining the total nebular \ha\ luminosity, the nebular physical
conditions. and chemical abundances.
We perform a detailed nebular empirical analysis of nine apertures extracted from 
a long-slit located to the west of HD\,37061 (east-west direction), obtaining the 
spatial distribution of the physical conditions and ionic abundances. For three of 
the analyzed elements (O, S, and N), we determine total abundances directly 
from observable ions (no ionization correction factors are needed). The comparison
of these abundances with those derived from the spectrum of the Orion nebula indicates
the importance of the atomic data and, specially in the case of M\,42, the considered 
ionization correction factors. 
}
{}
%
\maketitle
%
%
\section{Introduction}
\label{section1}
%
M\,43 (NGC1982) is an apparently spherical \hii\
region several arcmin to the northeast of the well-known Orion nebula. 
HD\,37061 (NU Ori, Par 2074, Brun 747), an early B-type star with broad 
lines located at the center of the nebula, is the main ionizing source
of M\,43. 
The spectral classifications found in the literature for HD\,37061 range 
between B0.5\,V and B1\,V, probably because of the 
difficulty in detecting the faint \ion{He}{ii}\,4541 line when using low 
resolution spectra and photographic plates. Up to three components have 
been identified within this stellar system \citep{Abt91, Pre99}. 
The primary star is known to be a spectroscopic binary (SB1) with an 
estimated stellar mass ratio $M_2$/$M_1$\,$\sim$\,0.19 \citep{Abt91}. 
Using bispectrum speckle 
interferometry, \cite{Pre99} found a third companion at $\sim$ 470 mas 
with a flux ratio in the K-band of \solu{0.03}{0.02}, implying a stellar 
mass ratio $M_3$/$M_{1,2}$\,$\sim$\,0.07.  Although the less massive 
components may affect the spectrum of the primary, they can be neglected 
in terms of ionization of the nebula.

M\,43 belongs to the same molecular complex as the Orion nebula \citep{Gou82}.
In particular, M\,43 is located at the northeast border of the extended Orion nebula (EON), 
a large elliptical structure surrounding the Huygens region (corresponding to the central, 
brighter part of the Orion nebula). Although the precise structure of the 
EON is still unknown, recent studies \citep{Ode09, Ode10} have associated 
the diffuse\footnote{The H$\alpha$ surface brightness in the EON is 
$\sim$~ two orders of magnitude lower than the brightest regions in M\,42.} and 
extended emission arising from this region with scattered light, produced 
by dust reflecting the stellar continuum emission from the 
Trapezium cluster stars and the nebular line emission from the Huygens region.
A dust lane defines the outer northern boundary of the EON and separates 
M\,43 from the Huygens region. As inferred from the optical polarization 
map of the M\,43 region obtained by \citet{Kha80}, this dust lane is not a 
foreground obscuration but a wall of dense material between M\,42 and M\,43.
This wall of material ensures that it is quite unlikely that ionizing 
light from the Trapezium cluster stars reaches the nebular material ionized
by HD\,37061; 
however, owing to the proximity of M\,43 to the {\em assumed} northern 
boundary of the EON, a scattered light contribution (similar 
to the one described above) may still affect affect the nebular emission received from M\,43.

Most studies of M\,43 have been performed at radio and far-infrared 
wavelengths. These studies were centered on the investigation of the global 
properties of the nebula \citep[i.e. geometry, \Te\ and \Ne,][]{Mil68, 
Thu78, Thr86, Sub92}, and the characteristics of the dust re-radiation 
inside the nebula \citep{Thu78, Smi87}. As for optical images, radio 
continuum observations indicate that M\,43 has a circular shape \citep{Sub92}. 
In addition, images at 60 $\mu$m published by \cite{Thr86}, and the kinematical
study by \cite{Han87} allow one to associate the nebula with a shell-shaped 
cavity on the side of a dense molecular ridge.

%
\begin{figure}[!t]
\centering
\includegraphics[width=8.5cm,angle=0]
{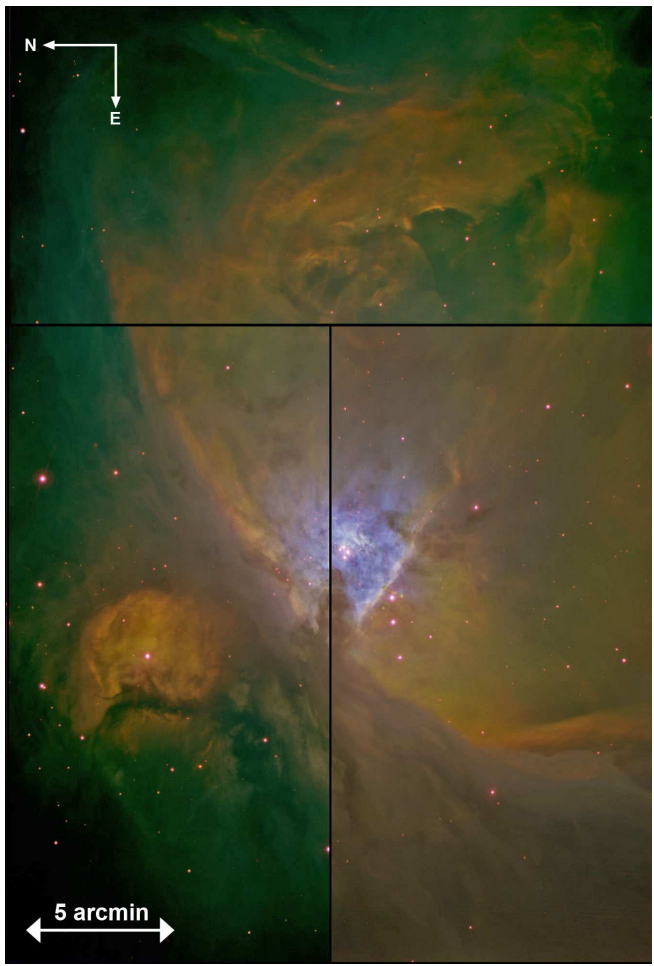}
\caption{Color composite (RGB) image of the M\,42+M\,43 region obtained from a
combination of narrow band images taken with the \wfc-\newton\ (only CCDs \#1, \#2, 
and \#4 are shown). The following color code was used: H$\alpha$ (red), 
[\ion{S}{ii}]\,$\lambda\lambda$6716+30 (green), [\ion{O}{iii}]\,$\lambda$5007 (blue). The field-of-view of the image is 
$\sim$\,22\,arcmin\,x\,33\,arcmin. M43 is the roundish nebulosity at the north-east 
of the center.}
\label{f1}
\end{figure}
%

Its proximity\footnote{The distance to HD\,37061 and M\,43 can be assumed to be similar to that of its 
companion nebula M\,42. Two different studies using
Very Long Baseline Array observations \citep{Men07, San07} determined the distance to 
some stellar objects in M\,42 to be \solu{414}{7}\,pc and 389$^{+24}_{-21}$\,pc, respectively.
These new measurements are $\sim$\,10\% lower than the 450 pc distance
often assumed for the Orion nebular cloud (ONC). We refer the reader to \citet{Men07} 
for a discussion of the different distances that have been determined
using different methods (viz. \citet{Jef07}: 392 pc, 
based on the statistical properties of rotation in pre-MS-stars; 
\citet{Gou82}: 480 pc, for a long time considered the canonical distance).}, 
relatively high surface brightness, simple geometry, and
isolated ionizing source make the Mairan's nebula an ideal object
to investigate several topics of interest in the field of \hii\ 
regions and massive stars. 
To this aim, we collected a set of high quality observations comprising 
the optical spectrum of HD\,37061, along with optical imaging and long-slit 
spatially resolved spectroscopy of M\,43. These observations were used to
perform a comprehensive study of M\,43 and its ionizing star, which is presented
in two parts. In this first paper, we (1) obtain the stellar parameters of HD\,37061, 
(2) investigate the presence of the scattered light in the region of M\,43 and 
its effect on the derived nebular properties, and (3) determine the physical conditions 
and nebular abundances in M\,43 obtained at various distances to the center of the nebula.
All this information is then used in a second paper (Sim\'on-D\'iaz et al., Paper II, in prep.) 
to construct a specialized photoionization model of the nebula using as many 
observational constraints as possible. The main drivers of this second part of our study 
are to (1) test the reliability of the ionizing spectral energy distributions (SEDs)
provided by the modern stellar atmosphere codes using nebular constraints, (2) 
investigate the effect of stellar pumping \citep{Fer99, Lur09} on the nebular 
emission arising from the inner part of this \hii\ region, and (3) test our 
understanding of the nebular energy budget temperature distribution.\\

%
\begin{figure*}[!t]
\centering
\includegraphics[width=17.cm,angle=0]
{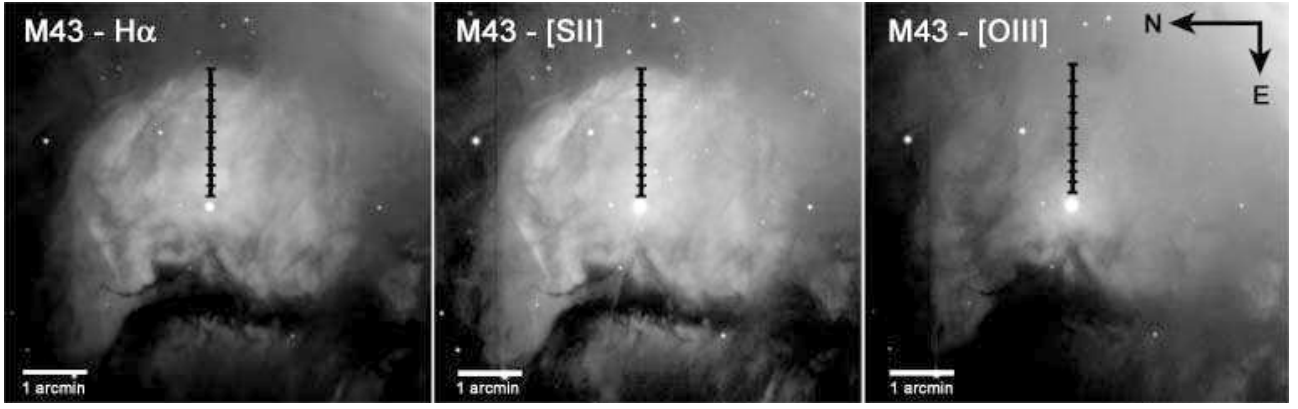}
\caption{\ion{H}{$\alpha$}, [\ion{S}{ii}]\,$\lambda\lambda$6716+30, and [\ion{O}{iii}]\,$\lambda$5007 \wfc-\newton\ images of 
M\,43. The size and position of the apertures extracted from the \isis-\herschel\ long-slit spectroscopic 
observations are also indicated.
}
\label{f2}
\end{figure*}
%
%

This paper is structured as follows. The observational data\,set is presented in Sect. 
\ref{section2}. A quantitative spectroscopic analysis of the optical 
spectrum of HD\,37061 is performed in Sect. \ref{section5}. The morphological 
characteristics of the nebula, 
along with its physical conditions and
nebular abundances are determined in Sects. \ref{section3} and \ref{section4} by analyzing
of the \ha, \hb, [\ion{O}{iii}], and [\ion{S}{ii}] narrow-band images, and the nebular
spectra, respectively. Both nebular analyses also allow us to find several indications
of an extended nebular emission, not directly related to M\,43, 
which must be adequately subtracted for a correct interpretation of the nebular images
and spectroscopy. Finally, we outline the main results of this study in Sect. \ref{section7}.

\section{The observational data\,set}\label{section2}
%

\subsection{Nebular imaging}
\label{section21}

%
%
\begin{table}[t!]
\begin{center}
\caption{\footnotesize Summary of the observations collected for the study.  
\label{t1}
}
\scriptsize{
\begin{tabular}{ccccc}    
\hline
\hline
\multicolumn{5}{c}{Nebular imaging}  \\
\hline
\noalign{\vskip2pt}
Instrument & Filter & \# & $\lambda_0$ (\AA) & FWHM  (\AA) \\
(Telescope) & &  & & \\
\noalign{\vskip1pt}
\hline
\wfc              &  \ha             & 197 & 6568 & 95   \\
(\newton)           &  \ha\ redsh.     & 228 & 6805 & 93   \\
(0.33 arcsec/pix) &  \hb\ narrow      & 225 & 4861 & 30   \\
                  &  \hb\ broad       & 224 & 4861 & 170  \\
                  &   [\ion{O}{iii}] & 196 & 5008 & 100  \\
                  &   [\ion{S}{ii}]  & 212 & 6725 & 80   \\
\hline
\multicolumn{5}{c}{Nebular spectroscopy}  \\
\hline
\noalign{\vskip2pt}
                  & Grid & $\lambda_0$ (\AA) & Spect. range (\AA) & $\Delta\lambda$ (\AA/pix)\\
\noalign{\vskip1pt}
\hline
\isis       & R600B & 4368 & 3386--5102 & 0.45 \\
(\herschel)   & R600R & 6718 & 6064--7585 & 0.49 \\
\hline
\multicolumn{5}{c}{Stellar spectroscopy}  \\
\hline
\noalign{\vskip2pt}
    & Grid & $\lambda_0$ (\AA) & Spect. range (\AA) & $\Delta\lambda$ (\AA/pix)\\
\noalign{\vskip1pt}
\hline
\ids             & H\,2400B & 4320 & 4060 - 4590 & 0.24 \\
(\newton)          & H\,2400B & 4800 & 4550 - 5070 & 0.24 \\
                 & H\,1800V & 6400 & 6090 - 6760 & 0.31 \\
\hline
\hline
\end{tabular}
\\
}
\end{center}
\end{table}

The Wide Field Camera\footnote{\tt{http://www.ing.iac.es/Astronomy/instruments/wfc/}} (\wfc) 
attached to the Isaac Newton Telescope (INT)
at the Roque de los Muchachos Observatory (La Palma, Spain) was 
used to obtain several narrow-band filter images of M\,43 in 2004 February 16, and
2006 October 21. The 
\wfc-\newton\ consists 
of four thinned EEV 2kx4k CCDs located at the primary focus of the telescope. 
Each CCD yields a field of view of 11.2' $\times$ 22.4', which results 
in a edge to edge limit of the mosaic (neglecting the $\sim$1 arcmin inter-chip 
spacing) of 34.2 arcmin. The pixel size is 13.5\,$\mu$m corresponding to 
0.33 arcsec/pixel. 

We centered M\,43 on chip\#4 of the \wfc, and obtained images of the 
region using the narrow-band filters  H$\beta$, [\ion{O}{iii}]\,$\lambda$5007, 
H$\alpha$, and [\ion{S}{ii}]\,$\lambda\lambda$6716+30.  
The log of the observations, along with the
characteristics of the various filters can be found in Table \ref{t1}.
While the [\ion{O}{iii}] and [\ion{S}{ii}] images were only obtained to 
provide some qualitative information about the nebula, we wished to produce 
pure emission-line, flux-calibrated images for H$\alpha$ and H$\beta$.
We hence obtained these images during a photometric night, and included 
the H$\alpha$ redshifted, and H$\beta$ broad filters in our set 
of observations, to correct the H$\alpha$ and H$\beta$ images 
for the adjacent continuum. The exposure times considered for the images in the
various filters were 4$\times$60\,s for \ha, 
4$\times$120\,s for \ha\ continuum, 3$\times$120\,s for \hb, 3$\times$180\,s for 
\hb\ continuum, 5$\times$12\,s for [\ion{O}{iii}], and 5$\times$12\,s for [\ion{S}{ii}].

The reduction of the images was performed following the standard procedures 
(trimming, bias subtraction, flat-fielding, and alignment) with 
\iraf\footnote{\iraf\ is distributed by NOAO which is operated by AURA Inc., under 
cooperative agreement with NSF}.  H$\alpha$ and H$\beta$ images were flux 
calibrated using observations of the spectrophotometric standard star BD\,+28$^{\circ}$4211 
\citep{Lan07} at different airmasses. We obtained the 
continuum-subtracted \ha\ and \hb\ images following the procedure described in 
\cite{Lop08}. 

The large field of view of the \wfc\ also allowed us to include M\,42 in the images.
A color-composite image of the M\,43+M\,42 region obtained from a combination
of the [\ion{O}{iii}] (blue), H$\alpha$ (green), and [\ion{S}{ii}] (red) images is 
presented in Fig.~\ref{f1}. Individual images in each one of these three filters, showing a square 
region of $\sim$\,6 arcmin centered in M\,43, are presented in Fig.~\ref{f2}.

\subsection{Stellar spectroscopy}
\label{section23}
%
%
\begin{figure}[t]
\centering
\includegraphics[height=8.5cm,angle=0]
{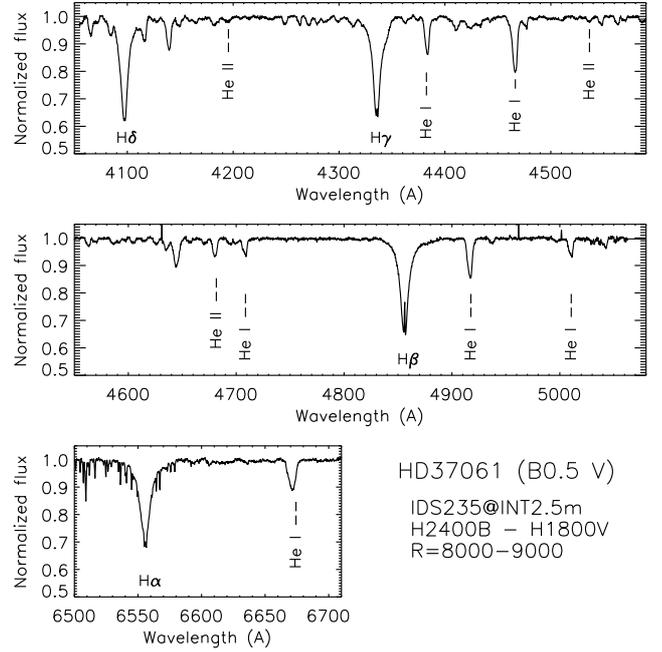} 
\caption{\ids-\newton\ optical spectrum of HD\,37061. The \ion{H}{i} and \ion{He}{i-ii} lines
used for the spectroscopic analysis of the star are indicated.}
\label{f4}
\end{figure}
%

The spectroscopic observations of HD\,37061 were carried 
out with the \newton\ on 2006 August 30 and September 5. The Intermediate 
Dispersion Spectrograph (\ids) was used with the 235\,mm camera and two 
different gratings (see Table \ref{t1}). We observed the spectral region between 
$\lambda\lambda$4000 and 5050\,\AA\ 
using the H2400B grating, which resulted in an effective resolving 
power R\,$\sim$\,7500 (equivalent to a 0.24\,\AA/pixel resolution 
and $\sim$\,2.6\,pixel \fwhm\ arc lines). For the \ha\ region 
($\sim$\,$\lambda\lambda$6090\,--\,6760\,\AA), the H1800V 
grating was used, resulting in a similar spectral resolution (0.31 \AA/pixel, 
R\,$\sim$\,8000). With these configurations, three exposures were 
needed to cover the whole range. We obtained spectra with exposure times of 180\,s and 70\,s for the blue (2)
and red (1) configurations, respectively, and obtained two spectra per spectral
range. A large number of flat fields and 
arcs for the data reduction process were obtained.

The reduction and normalization of the spectra was made following 
standard techniques, with \iraf\ and our own software developed in \idl. 
The signal-to-noise ratio of the reduced spectra is about 300\,--\,400, depending on the 
spectral range. Fig.~\ref{f4} shows the global stellar spectrum, along with the \ion{H}{i}
and \ion{He}{i-ii} lines used for the quantitative spectroscopic analysis.

\subsection{Nebular spectroscopy}
\label{section22}

%
\begin{figure}[t]
\centering
\includegraphics[width=8.5cm,angle=0]
{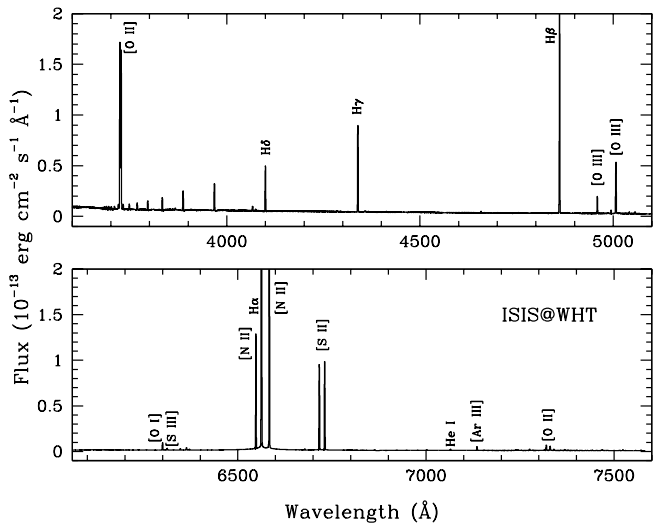}
\caption{\isis-\herschel\ optical spectrum of M\,43 (aperture \#5). The \ion{H}{i} Balmer
lines, along with \ion{He}{i} and other metal lines used in our study, are indicated.}
\label{f3}
\end{figure}
%

A long-slit, intermediate-resolution spectrum of M\,43 was obtained on 
2006 December 23 with the Intermediate dispersion Spectrograph and Imaging 
System (\isis) spectrograph attached to the 4.2m William Herschel 
Telescope (\herschel) at Roque de los Muchachos Observatory (La Palma, Spain). 
The 3.7$'$\,$\times$\,1.02$''$ slit
was located to the west of HD\,37061, along the E-W direction (PA=90$^{\circ}$, 
see Fig.~\ref{f2}). Two different CCDs were used at the blue and red arms of 
the spectrograph: an EEV CCD with a configuration 4096 $\times$ 2048 pixels at 
13.5 $\mu$m, and a RedPlus CCD with 4096 $\times$ 2048 pixels 
at 15.0 $\mu$m, respectively. The dichroic used to separate the blue and red 
beams was centered on 5400 \AA.  The gratings R600B and R600R were used for the blue 
and red observations, respectively (see log of the observations in Table \ref{t1}). 
These gratings give a reciprocal dispersion of 33 \AA\ mm$^{-1}$ in both 
cases, and effective spectral resolutions of 2.2 and 2.0 \AA, respectively. 
The blue spectra cover from $\lambda\lambda$3386 to 5102 \AA\ and the red ones 
from $\lambda\lambda$6064 to 7585 \AA. The spatial scales are 0.20$''$ pixel$^{-1}$ 
and 0.22$''$ pixel$^{-1}$, respectively. The seeing during the 
observations was between $\sim$0.5$''$ and $\sim$0.8$''$. The exposure times were 
3$\times$300\,s, in both the blue and red observations (obtained at the same time).

The spectra were wavelength calibrated with a CuNe+CuAr lamp. The correction 
for atmospheric extinction was performed using the average curve for continuous 
atmospheric extinction at Roque de los Muchachos Observatory. The absolute 
flux calibration was achieved by observations of the standard stars 
HD\,19445, H\,600, and HD\,84937. We used the standard \iraf\ TWODSPEC reduction 
package to perform bias correction, flat-fielding, cosmic-ray rejection, wavelength, 
and flux calibration. Fig.~\ref{f3} shows an illustrative example of the 
global appearance of our nebular spectroscopic observations, also indicating 
the main nebular lines used in this study.

\section{Quantitative spectroscopic analysis of HD\,37061}
\label{section5}
%
The stellar parameters of the star were derived by visually 
comparing the observed optical \ion{H}{i}, \ion{He}{i}, and \ion{He}{ii} 
line profiles with the synthetic ones resulting from the stellar atmosphere 
code \fastwind\ \citep{San97, Pul05}, which is an established technique. 
Similar analyses, along with some notes on 
the methodology, can be found in \cite{Her02}, \cite{Rep04}, \cite{Sim06}, 
and references therein.

To this aim, we constructed a grid of \fastwind\ models with \Teff, and \grav, 
ranging from 28000 to 33000 K (500 K steps), and from 3.9 to 4.3 dex (0.1 dex steps), respectively. The microturbulence,
the helium abundance, and the wind-strength parameter\footnote{$Q\,=\,\dot{M}/(R\,v_{\infty})^{1.5}$.} 
were fixed to characteristic values for an early B dwarf star (\micro\,=\,5 \kms,
$\epsilon$\,=\,0.09, and log\,Q\,=\,-15, respectively) and metal abundances were assumed to be solar 
\citep[following the set of abundances by][]{Asp09}. 

This technique requires the projected rotational velocity (\vsini) of the star to be
previously established. All metal lines in the spectrum of HD\,37061 are blended or 
very shallow because of the large \vsini\ of the star; therefore, we decided to apply the 
Fourier method (c.f. \citeauthor{Gra76} \citeyear{Gra76}; see also \citeauthor{Sim07} 
\citeyear{Sim07} for a recent application to OB-type stars) to the \ion{He}{i} 
lines, and obtained a \vsini\,$\sim$\,200 \kms.

%
%
\begin{figure}[t!]
\centering
\includegraphics[width=8.5cm,angle=0]{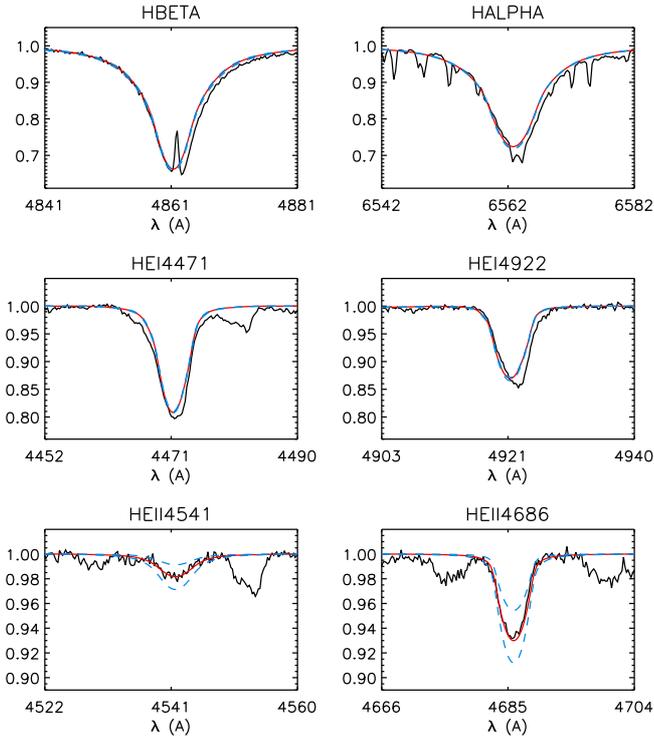}
\caption{Fit of FASTWIND \ion{H}{i} and \ion{He}{i-ii} synthetic line profiles to 
         the observed ones (red solid line, \Teff=31000 K, \grav=4.2 dex). Blue
         dashed lines illustrate the effect of a variation of $\pm$1000 K in the
         FASTWIND models. Note that the red wings of the \ion{H}{i} and \ion{He}{i}
         lines are affected by the cool companion.}
\label{f_fit}
\end{figure}

%
\begin{table}[t!]
\begin{center}
\caption{\footnotesize Stellar parameters derived through the 
         \fastwind\ analysis of the optical spectrum of HD\,37061.}
\label{t5}
\begin{tabular}{l c | l c}    
\hline
\hline
\noalign{\smallskip}
\Teff (K)            & \solu{31000}{500} & R (R$_{\odot}$)  & \solu{5.7}{0.8}    \\
\grav (dex)          & \solu{4.2}{0.1}    & logL/L$_{\odot}$ & \solu{4.42}{0.12} \\
$\epsilon$(He)       & 0.09 (assumed)     & M (M$_{\odot}$)  & \solu{19}{7}       \\
log\,Q               & -15 (assumed)      & log\,Q(H$^0$)    & \solu{47.2}{0.2} \\
\noalign{\smallskip}
\hline
\hline
\\
\end{tabular}
\end{center}
\end{table}

The synthetic \ion{H}{i} and \ion{He}{i-ii} line profiles resulting from the models were then 
convolved with the corresponding instrumental and rotational profiles and
compared with the observations.
The best fit was found for \Teff\,=\,31000 K and \grav\,=\,4.2 dex (see 
Fig. \ref{f_fit}). Given the quality of the spectrum and the sensitivity of 
the used lines to the variation in \Teff\ and \grav\ for this range of 
stellar parameters, an accuracy of 500 K and 0.1 dex, respectively, 
could be achieved. As an example, in Fig. \ref{f_fit} we show the 
effect of a variation of $\pm$1000 K in the FASTWIND models. For this
range of stellar parameters, the \ion{He}{ii} lines are decisive in constraining
the \Teff.

We note the poor quality of the fits in the red wings of the \ion{H}{i} and \ion{He}{i}
lines (but not in the \ion{He}{ii} lines). These lines are likely affected by the spectrum
of the cooler secondary spectroscopic component (see Sect. \ref{section1}).

Once the stellar parameters have been determined, the stellar
radius can be derived from the absolute visual magnitude ($M_{\rm V}$) and the 
synthetic spectral energy distribution provided by the stellar atmosphere model. 
By using $m_{\rm V}$=6.84 and $A_{\rm V}$=2.09
\citep{Hil97}, and adopting a distance to the star of 400$\pm$50 pc, we obtained
an $M_{\rm V}$\,=\,-3.3$\pm$0.3, quite consistent with the spectral type of HD\,37061. 
The resulting stellar parameters, along with the total number of ionizing photons
are summarized in Table \ref{t5}.

\section{Qualitative and quantitative analysis of the nebular images}
\label{section3}
%
%
Fig.~\ref{f1} shows a color composite image of the M\,42+M\,43 region.
The object under study, M\,43, appears as a roundish \hii\ region
centered around HD\,37061, and well-separated from the Orion nebula (M\,42)
by a dust lane (known as the northeast dark lane). The nebula has a 
diameter of $\sim$4.5\,arcmin ($\sim$\,0.65\,pc at a distance of 400\,pc), and is
crossed by a dark lane oriented N-S in the eastern side (known as the M43 dark lane). 
This dark cloud is located in front of the nebula, blocking the nebular 
light coming from behind.

Fig.~\ref{f2} shows a closer view of the nebula, where three images of M\,43, taken
through filters isolating H$\alpha$ (+[\ion{N}{ii}]\,$\lambda\lambda$6548+84),
[\ion{S}{ii}]\,$\lambda\lambda$6716+30, and [\ion{O}{iii}]\,$\lambda$5007, are presented
separately. The images show diffuse and extended emission
beyond the limits of the nebula (mainly to the south and west). This 
emission can be more clearly seen in the [\ion{O}{iii}] image. If the entire nebula were spherical 
with a density of 500 cm$^{-3}$,
the size of the O$^{2+}$ region would be $\sim$\,25\,\% of the total
size of the nebula\footnote{Computed from a simple spherical \cloudy\ model using the
ionizing spectral energy distribution from a FASTWIND model with \Teff\,=\,31000\,K, and \grav\,=\,4.2.}; 
however, Fig.~\ref{f2} shows that the [\ion{O}{iii}] emission originates in a more extended 
region (i.e. is unlikely to be related\footnote{This extended emission is more likely to be associated 
with the scattered nebular light arising from the EON, described by \citet{Ode09} and \citet{Ode10}.
See a more detailed discussion about this possibility in Sect. \ref{section49}.}
to nebular material ionized by HD\,37061). 

\begin{figure}[!t]
\centering
\includegraphics[width=8cm,angle=0]
{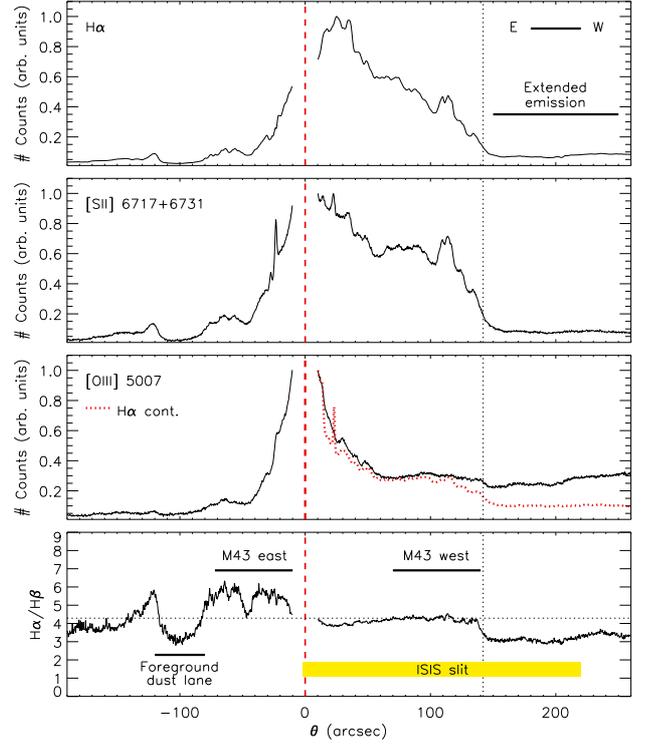}
\caption{Three upper panels: Spatial distribution of \ha, [\ion{S}{ii}]\,$\lambda\lambda$6716+30, 
and [\ion{O}{iii}]\,$\lambda$5007 emission in a line passing by HD\,37061 in the west-east 
direction (see Fig.~\ref{f2}), obtained from the corresponding WFC-INT images. 
The dashed red line indicates the position of HD\,37061. The vertical dotted 
line indicate the western limit of M\,43. Bottom panel:
\ha/\hb\ spatial distribution in the same line direction. Some dust features
outlined by \citet{Smi87} and the position and size of the ISIS-WHT slit are 
indicated. Horizontal dotted lines show the \ha/\hb\ value in the M43 west
region.}
\label{f5}
\end{figure}

To more clearly illustrate this, we plot in Fig.~\ref{f5} (three upper panels) the spatial distribution of
\ha, [\ion{S}{ii}], and [\ion{O}{iii}] emission in a line passing by HD\,37061 along 
the east-west direction, obtained from the corresponding \wfc-\newton\ images. Since
we are only interested in the relative spatial changes, the three distributions
are normalized to their maximum value. In all cases, a faint emission,
not expected to be related to M\,43, is clearly detected beyond the limits
of the nebula. In the [\ion{O}{iii}] panel, we also superimpose the spatial distribution
derived from the \ha-continuum image to show that the extended emission observed in the [\ion{O}{iii}]
image is not associated with the nebular continuum.

%
%
%
\subsection{Extinction}
\label{section31}
%
The flux-calibrated, continuum-subtracted H$\alpha$ and H$\beta$
images (i.e. pure emission) were used to obtain the H$\alpha$/H$\beta$ flux ratio
distribution along the diameter in the east-west direction passing
by the ionizing star of M\,43 (see bottom panel in Fig.~\ref{f5}). 
Some interesting information about the extinction in the studied region can be 
extracted by inspecting of this figure: 
\begin{itemize}
 \item[a)]{There is a jump in the H$\alpha$/H$\beta$ flux ratio
140 arcsec west of HD37061 (i.e. western boundary of M\,43), indicating
a clear difference in the type of emission arising from outside the limits of
the nebula. We note that the lower H$\alpha$/H$\beta$ ratio 
found in the ``extended emission'' region agrees with the hypothesis
that this emission is associated with scattered light (see also Sect. \ref{section49}), 
and is not necessarily an indication of a lower extinction in this region.}
\item[b)]{The H$\alpha$/H$\beta$ ratio in the western region of M\,43 is fairly 
constant, though there is a small decrease close to the star. On the other hand, 
this ratio is larger in the eastern part.}
\item[c)]{The M\,43 dark lane location (labeled as a foreground dust lane in 
Fig. \ref{f5}) is clearly represented in the H$\alpha$/H$\beta$ distribution.} 
\end{itemize}

\citet{Smi87} presented a general view of the spatial distribution of dust in the
M\,43 region. They found 
three regions where dust reradiation in 60 $\mu$m is concentrated (see a schematic 
map of dust features in their Fig 1.c). The first one is associated with the M\,43 dark lane 
(they called it the foreground dust lane), the second one (M\,43 east) is located 
to the east of HD\,37061, between the ionizing star and the M\,43 dark lane, 
and a third one (M\,43 front\,+\,M\,43 west) follows the border of the 
nebula from the north to the west. This third feature of dust emission is 
separated $\sim$\,60-70 arcsec from the star, leaving a ``dust-empty'' space
in-between. We indicate in the bottom panel of Fig. \ref{f5} the location 
of these dust features using the same nomenclature as \citeauthor{Smi87}

The observed behavior of the H$\alpha$/H$\beta$ flux ratio within M\,43 along the 
west-east direction is perfectly correlated with the dust distribution presented
by \citet{Smi87}. It is remarkable that, in contrast to expectations, the amount of
extinction indicated by the H$\alpha$/H$\beta$ flux ratio in the foreground
dust lane is very low. The explanation is simple. This dark feature obscures
the emission from M\,43, hence the observed light originates in the material 
located in front. This emission follows the trend indicated by the region marked as
``extended emission''.

%
\subsection{Total nebular H$\alpha$ luminosity and surface brightness}
\label{section32}

%
%
%
\begin{figure}[t!]
\centering
\includegraphics[height=8.cm,angle=90]{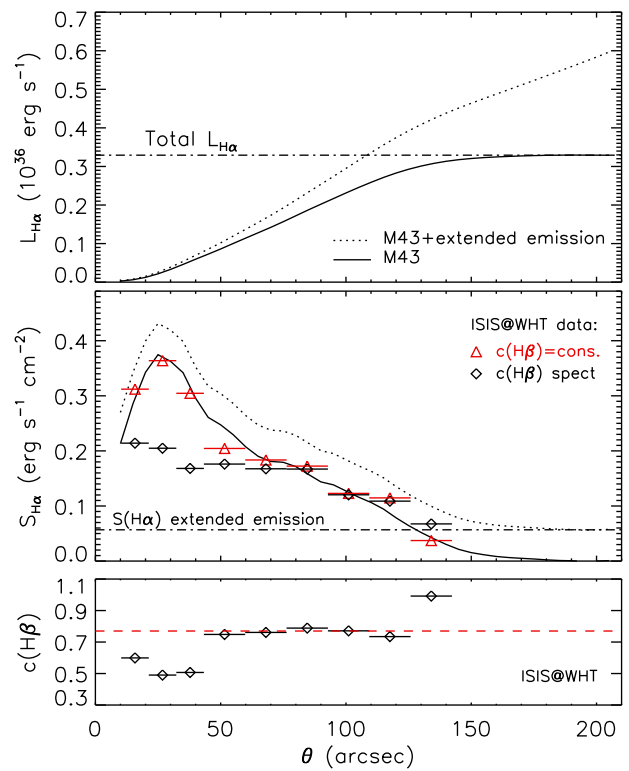}
\caption{Top and middle panels: integrated \ha\ luminosity profile and extinction 
corrected \ha\ surface brightness profile, respectively, obtained from the nebular image. 
Bottom panel: reddening coefficient obtained from the empirical analysis of the nebular
spectra (see Sect.~\ref{section4}). 
In the upper panels we compare the L(\ha) and S(\ha) profiles of the original 
images (dotted lines) with those of images  corrected for the contribution of the 
extended emission (solid lines). Some data obtained from
the ISIS-WHT spectroscopic observations are also included as triangles and
diamonds (see text for explanation).}
\label{f6}
\end{figure}

To derive the total nebular H$\alpha$ luminosity we proceeded as
follows. First, the H$\alpha$ image was corrected for the effects of distance, 
extinction, and [\ion{N}{ii}] contamination\footnote{Note that what we
refer to as a ''H$\alpha$ image'' is actually a ''H$\alpha$+[\ion{N}{ii}]\,$\lambda\lambda$6548+84''
image.}. Denoting with $F_{\rm H\alpha}$ the nebular flux received at Earth, 
the corresponding extinction-corrected H$\alpha$ luminosity can be derived from:
\begin{equation}
L_{\rm H\alpha}=4\pi d^2 F_{\rm H\alpha} 10^{c_{\rm H\beta}\,{f(\rm H\alpha)}},
\end{equation}
where $d$ is the distance to the nebula, $c$({\hb}) is the reddening coefficient, and $f(\rm H\alpha)$ is
the value of the extinction function in H$\alpha$ relative to H$\beta$.

The analysis of the H$\alpha$/H$\beta$ image (Section \ref{section31})
showed that the extinction is not constant across the nebula. The 
optimal strategy to follow would be to 
obtain $c$({\hb}) for every pixel from the H$\alpha$/H$\beta$ images, 
correct the H$\alpha$ image, and then integrate the whole nebula to obtain the 
total L(H$\alpha$). 
However, when we attempted to follow this strategy we found that 
it introduces many sources of uncertainty. In addition, the nebular region
hidden by the dark lane cannot be corrected in this way. We therefore decided to 
follow a different approach, taking into account what we learned from the inspection
of the nebular images, and making use of the information extracted from the 
spectroscopic observations (see Sect.~\ref{section4}). We decided to integrate only the 
west quadrant of the nebula, and multiply the resulting value by four (i.e.
we assume the nebula is symmetric). Extinction in this quadrant
is lower and more constant than in the other three. Furthermore, we have information
about $c$({\hb}) ($\sim$\,0.76) and the nitrogen contribution to the \ha\ image 
([\ion{N}{ii}]\,$\lambda\lambda$6548+84/H$\alpha$\,$\sim$\,0.59) for this region from 
the ISIS-WHT spectroscopy (see Sect.~\ref{section4}).

Once the distance, extinction, and nitrogen corrections had been applied to the continuum-subtracted, 
flux-calibrated \ha\ image,
we obtained the L(H$\alpha$) inside circles of increasing sizes\footnote{Actually, we 
integrated the mentioned quadrant and multiply the obtained value by four.} (centered 
on the star), and the surface brightness distribution within concentric rings of 
increasing radii. Fig.~\ref{f6} shows the measured quantities as a function
of distance to the star. Dotted lines show the initial determinations where the images
were not corrected for the extended emission contribution. The \ha\ surface 
brightness distribution again shows
that outside the limits of the nebula ($\theta\,\ge$\,160 arcsec) there remains a non-negligible, 
more or less constant emission. This implies that L(H$\alpha$) increases continuously, even
outside the nebula (top panel in Fig.~\ref{f6}). We therefore assumed a 
constant\footnote{We note 
that the H$\alpha$ surface brightness of the extended emission is not necessarily constant 
across the nebula; however, its variation is negligible compared to the H$\alpha$ 
emission from M\,43, hence can be considered to be constant for the sake of simplicity.}
contribution of the extended emission to the surface brightness ($\sim$\,0.06\,erg\,s$^{-1}$\,cm$^{-2}$, see
middle panel of Fig. \ref{f6}) and subtracted it from the \ha\ 
image. The resulting L(H$\alpha$) and S(H$\alpha$) distributions, calculated from the
corrected images in the same way as described above, are indicated as solid lines in the top and middle panels
of Fig.~\ref{f6}. These quantities behave as expected, i.e., the surface brightness
becomes zero, and L(H$\alpha$) remains constant for R$>$R$_{\rm neb}$. The resulting
total \ha\ luminosity is (3.3$\pm$1.1)\,x\,10$^{\rm 35}$ erg\,s$^{\rm -1}$, where the 
associated uncertainty was calculated by taking into account uncertainties of
12.5\%, 10\%, and 20\% in distance, $F$(\ha), and $c$({\hb}), respectively.

The bottom panel of Fig.~\ref{f6} shows the $c$({\hb}) values obtained from the
spectroscopic analysis of the 9 apertures extracted from the long-slit ISIS-WHT
observations (see Sect.~\ref{section42}). As for the \ha/\hb\
images (bottom panel in Fig.~\ref{f5}), we found that $c$({\hb})
is fairly constant for $\theta\,>$\,50\,arcsec, but is somewhat smaller in the
inner region (close to the star). We also include some information obtained from
the spectroscopy in the middle panel of Fig.~\ref{f6} (surface brightness profile).
Red triangles correspond to the calculated S(\ha) in each aperture assuming a constant
$c$({\hb}) of 0.76. The derived values follow the
S(\ha) profile obtained from the \ha\ image. The middle panel in Fig.~\ref{f6} also
shows the corresponding S(\ha) values computed from the actual $c$({\hb})
values derived from the spectroscopic analysis (black diamonds). 
These values deviate from the constant extinction-corrected values in the inner part 
of the nebula. We can conclude that the increase 
in S(H$\alpha$) in the inner part of the nebula is not necessarily caused by an
increase in the emission measure or a deviation from the spherical geometry, but is more
likely to be an effect of a non-constant extinction or stellar pumping effects on H recombination
lines \citep[see][]{Fer99, Lur09}. 
The total L(\ha) we derived above must hence be corrected for this
effect, resulting in (3.0$\pm$1.1) x 10$^{\rm 35}$ erg\,s$^{\rm -1}$ (about 10\% 
below the previously derived value).

In Sect. \ref{section5}, we derived the total number of ionizing photons emitted by
the central star (Q(H$^0$)\,=\,10$^{47.2\,\pm\,0.2}$ photons\,s$^{-1}$). This
would imply a maximum nebular H$\alpha$ luminosity of (2.5$\pm$1.0)\,x\,10$^{35}$
erg\,s$^{-1}$, a value that is in quite good agreement with the total H$\alpha$ 
luminosity obtained from the analysis of the nebular images. This result is
compatible to first order with M\,43 being a (mostly) ionization-bounded 
nebula in which dust does not absorb any significant fraction of the Lyman 
continuum photons.

We note that the stellar and nebular results would be incompatible if we
considered the total nebular H$\alpha$ luminosity obtained from the images
that are not corrected for the extended emission contribution. The total number of ionizing
photons necessary to explain the derived nebular H$\alpha$ luminosity would
be at least double the value obtained from the analysis of the star.

%
\section{Empirical analysis of the nebular spectra}
\label{section4}
%
\subsection{More indications of extended nebular emission}
\label{section41}

Fig.~\ref{f7} shows the spatial distribution of the nebular emission in four selected
lines (\ha, \foii\,$\lambda\lambda$3720+30, \ion{He}{i}\,$\lambda$6678, and
\foiii\,$\lambda$5007) and their respective adjacent nebular continua. These 
distributions were obtained
from the ISIS@WHT spectroscopic observations. The colored lines represent the 
measured line fluxes (blue), the adjacent continuum emission (red), and the continuum-subtracted 
line fluxes (black). These figures display non-negligible emission beyond the 
limits of the nebula ($\theta\,>\,$150 arcsec), that is not 
associated with the continuum emission. In 
most cases, this external emission remains fairly constant in the studied
region (e.g. \ha, \foii\,$\lambda\lambda$3720+30, \ion{He}{i}\,$\lambda$6678, as 
well as \fnii, \fsii, and \fsiii\ lines). However, in some cases (e.g. \foiii), 
this emission increases with increasing distance from the star. 

\subsection{Aperture selection and line flux measurements}
\label{section42}

%
\begin{figure}[!t]
\centering
\includegraphics[width=8.5cm,angle=0]
{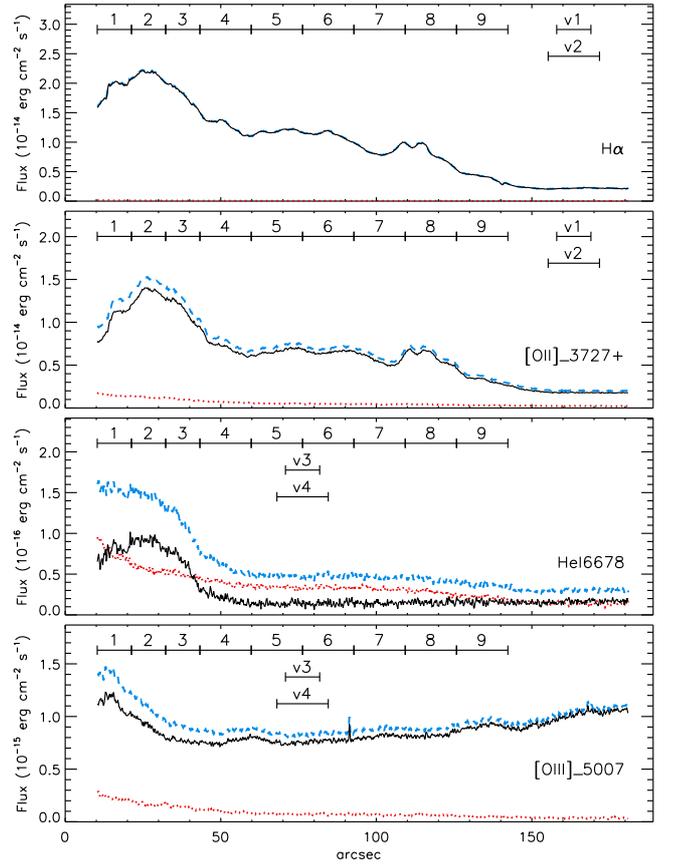}
\caption{Spatial distribution of the nebular emission in four selected
nebular lines (\ha, \foii\,$\lambda\lambda$3720+30, \ion{He}{i}\,$\lambda$6678, and
\foiii\,$\lambda$5007) and their respective adjacent nebular continua, obtained from the ISIS-WHT 
spectroscopic observations. The colored lines represent the measured line fluxes (blue), 
the adjacent continuum emission (red), and the continuum-subtracted line fluxes (black).
The apertures used to extract nebular spectra at different distances from the central star
are also indicated (see Sect.~\ref{section42}). Apertures labeled as v1, v2, v3, and v4, are 
those used to correct the other nine apertures for the contamination by the extended emission component.}
\label{f7}
\end{figure}
%

We obtained one-dimensional (1D) spectra of regions of the nebula at different distances
from the central star by dividing the long-slit used for the \isis-\herschel\ nebular 
observations into nine small apertures within the limits of the nebula ($\theta<$150 arcsec)
with the aim of obtaining: a) the radial distribution of the physical 
properties of the nebula, b) information about nebular ionic and total abundances, 
and c) the radial distribution of critical nebular line ratios used as observational 
constraints on photoionization models of the nebula. 

The size and position of the apertures are summarized in Table \ref{t2} 
and shown in Figs. \ref{f2} and \ref{f7}. It is crucial to correct these spectra 
for the contribution of the extended emission component to obtain 
information about M\,43 itself. To achieve this aim, spectra from two extra-apertures outside
the nebula were extracted to correct the line flux measurements 
of the apertures for the extended emission component. The size of the extra-apertures 
was selected to be the same as the corresponding apertures inside the nebula (11 and 16.5 arcsec 
for the three innermost and the other outer apertures, respectively). Two different
locations of these extra-apertures were considered depending on whether the considered
nebular emission line had been emitted in the whole nebula (i.e. {\hi}, {\foii}, {\fnii}, {\fsii} 
and {\fsiii} lines; apertures v1 and v2 in Fig,~\ref{f7}), or only in the inner region 
(i.e. {\hei}, {\foiii}, and {\fariii} lines; apertures v3 and v4 in Fig.~\ref{f7}).
For the latter set of lines, it is impossible to use the same
apertures (v1 and v2) as for the other lines, since the extended emission in these
cases is not constant, but increases far away from the star. We hence considered
two apertures as being close to the region where we were confident that the contribution of 
the nebular emission associated with M\,43 had vanished. We expect the spectrum from
these apertures to represent the properties of the extended emission close to the
region where M\,43 is emitting in the given nebular line.

The extraction of 1D spectra from each aperture was performed using the \iraf\ 
task $apall$. The same zone and spatial coverage was considered in the blue 
and red spectroscopic ranges. 

We detected {\hi} and {\hei} optical recombination lines, along with 
collisionally excited lines (CELs) of several ions, such as {\foi}, {\foii}, 
{\foiii}, {\fnii}, {\fsii}, {\fsiii}, {\fneiii}, and {\fariii}. Line fluxes were 
measured using the SPLOT routine of the \iraf\ package by integrating all the flux 
included in the line profile between two given limits and over a local continuum 
estimated by eye. Although we detected two {\fneiii} lines, these were unsuitable 
for analysis because they were extremely faint (in the case of $\lambda$3869) or 
were severely blended with {\hi}$\lambda$3967. 
In the case of {\foii} $\lambda$7330 \AA , we corrected 
the flux for the overlapping telluric emission line 
OH 8--3 P2 $\lambda$7329.148 \AA\ by estimating
its intensity from the observed OH 8--3 P2 $\lambda$7329.148 \AA /OH 8--3 
P1 $\lambda$7340.885 \AA\ line ratio in standard star exposures.

Each emission line in the spectra from the nine apertures was corrected for the 
extended emission component by subtracting the corresponding emission measured in the appropriate 
extra-aperture (v1 to v4). Line intensities were then normalized to a particular {\hi} 
recombination line present in each wavelength interval ({\hb} and {\ha} for 
the blue and red spectra, respectively). 
To produce a final homogeneous set of line intensity ratios, the red spectra were then
rescaled to \hb\ using the theoretical \ha/\hb\ ratio I(\ha)/I(\hb)\,=\,2.92 obtained 
for the physical conditions of $T_e$\,=\,7500 K and $n_e$\,=\,500 cm$^{-3}$ (see Sect. \ref{section45}).

%
\begin{table*}
\centering \caption{Line intensities corrected from foreground emission and extinction (\hb\,=100), and
results from the empirical analysis of the nebular spectra corresponding to the nine extracted apertures$^{(1)}$. }
\label{t2}
\scriptsize
\begin{tabular}{c@{\hspace{2.8mm}}c@{\hspace{2.8mm}}c@{\hspace{2.8mm}}c@{\hspace{2.8mm}}c@{\hspace{2.8mm}}c@{\hspace{2.8mm}}c@{\hspace{2.8mm}}c@{\hspace{2.8mm}}c@{\hspace{2.8mm}}c@{\hspace{2.8mm}}c@{\hspace{2.8mm}}c@{\hspace{2.8mm}}}
\noalign{\hrule} 
\noalign{\vskip3pt}
   &    &  & \multicolumn{9}{c}{Aperture} \\
\cline{4-12}
\noalign{\vskip3pt}
   &    &  &  A1  & A2  & A3  & A4  & A5  & A6  & A7 & A8  & A9 \\
\cline{4-12}
\noalign{\vskip3pt}
\multicolumn{3}{r}{Center position (arcsec)} &  15.80 & 26.80 & 37.80 & 51.55 & 68.05 & 84.55 & 101.05 & 117.55 & 134.05 \\
\multicolumn{3}{r}{Size (arcsec)}&  11.0  & 11.0  & 11.0  & 16.5  & 16.5  & 16.5  & 16.5  & 16.5  &  16.5  \\
\noalign{\vskip3pt}
\noalign{\hrule} 
\noalign{\vskip3pt}
$\lambda$ (\AA) & Ion & Mult. & \multicolumn{7}{c}{$I$($\lambda$)/$I$(H$\beta$)}   \\
\noalign{\vskip3pt} \noalign{\hrule} \noalign{\vskip3pt}
3726.03	      & {\foii}       	& 1F	       &    93.5$\pm$1.8&	 99.7$\pm$2.0  &       106$\pm$2     &     104$\pm$2      &      104$\pm$2    &      108$\pm$2     &     117$\pm$2     &      129$\pm$3    & 	   150$\pm$3    \\
3728.82	      & {\foii}       	& 1F	       &    84.2$\pm$1.7&	 91.5$\pm$1.8  &      98.1$\pm$2.0   &    99.5$\pm$2.0    &     98.9$\pm$2.0  &      103$\pm$2     &     114$\pm$2     &      120$\pm$2    & 	   143$\pm$3    \\
3835.39	      & {\hi}	      	& H9$^{(2)}$	&    7.50$\pm$0.75&	 7.94$\pm$0.79 &      8.28$\pm$0.83  & 	  7.91$\pm$0.79   &     7.83$\pm$0.78 &     7.57$\pm$0.76  &     7.98$\pm$0.80 &     7.37$\pm$0.74 & 	  7.42$\pm$0.74 \\
4068.60       & {\fsii}  	& 1F   	       &     1.58$\pm$0.32&	1.80$\pm$0.36  &      1.86$\pm$0.37  &     2.43$\pm$0.49  &	2.70$\pm$0.54 &     3.30$\pm$0.66  &	 3.84$\pm$0.77 &     4.77$\pm$0.95 &	  5.63$\pm$0.56 \\
4076.35       & {\fsii}  	& 1F   	       &     0.64$\pm$0.19&	0.67$\pm$0.20  &      0.76$\pm$0.23  &     0.74$\pm$0.22  &	0.71$\pm$0.21 &     1.00$\pm$0.30  &	 1.16$\pm$0.23 &     1.29$\pm$0.26 &	  1.24$\pm$0.25 \\
4101.74	      & {\hi}    	& H$\delta$    &     26.2$\pm$1.3 &     26.2$\pm$1.3   &      26.0$\pm$1.3   &     25.7$\pm$1.3   &     25.6$\pm$1.3  &     25.6$\pm$1.3   &	 25.7$\pm$1.3  &     25.6$\pm$1.3 &	  25.8$\pm$1.3  \\
4340.47	      & {\hi}    	& H$\gamma$    &     46.2$\pm$2.3 &     46.2$\pm$2.3   &      46.5$\pm$2.3   &     46.9$\pm$2.3   &     46.9$\pm$2.4  &     47.0$\pm$2.4   &	 46.8$\pm$2.3  &     46.9$\pm$2.4 &	  46.8$\pm$2.3  \\
4471.09	      & {\hei}    	& 14	       &     1.09$\pm$0.22&	1.35$\pm$0.27  &      1.02$\pm$0.20  &     0.19$\pm$0.08  &	    ---     & 	---      &	     ---     &	 ---     &	      ---	\\
4861.33	      & {\hi}	    	& H$\beta$     &      100$\pm$2   &      100$\pm$2     &      100$\pm$2      &      100$\pm$2     &      100$\pm$2    &      100$\pm$2     &      100$\pm$2    &      100$\pm$2    &       100$\pm$2    \\
4958.91	      & {\foiii}  	& 1F	       &     2.52$\pm$0.50&	0.98$\pm$0.29  &      0.15$\pm$0.06  &         ---      &	    ---     & 	---      &	 0.93$\pm$0.28 &     0.88$\pm$0.26 &	  7.59$\pm$0.76 \\
5006.94	      & {\foiii}  	& 1F	       &     7.86$\pm$0.79&	2.78$\pm$0.56  &      0.25$\pm$0.10  &         ---      &	0.35$\pm$0.14 &     0.13$\pm$0.05  &	 2.78$\pm$0.56 &     3.06$\pm$0.61 &	 25.42$\pm$1.27 \\
6300.30	      & {\foi}		& 1F   	       &     0.64$\pm$0.03&	0.64$\pm$0.03  &      0.60$\pm$0.03  &     0.86$\pm$0.04  &	1.05$\pm$0.05 &     1.30$\pm$0.06  &	 2.47$\pm$0.12 &     3.50$\pm$0.17 &	  5.39$\pm$0.27 \\
6312.10       & {\fsiii}  	& 3F   	       &     0.80$\pm$0.04&	0.70$\pm$0.03  &      0.69$\pm$0.03  &     0.58$\pm$0.03  &	0.47$\pm$0.02 &     0.42$\pm$0.02  &	 0.45$\pm$0.02 &     0.38$\pm$0.02 &	  0.36$\pm$0.02 \\
6548.03       & {\fnii}  	& 1F   	       &    36.31$\pm$1.82&     38.4$\pm$1.9   &      41.3$\pm$2.1   &     41.8$\pm$2.1   &     42.0$\pm$2.1  &     44.2$\pm$2.2   &	 45.7$\pm$2.3  &     50.2$\pm$1.0  &	  54.1$\pm$1.1  \\
6562.82       & {\hi}  		& H$\alpha$    &   292.00$\pm$5.84&      292$\pm$6     &       292$\pm$6     &      292$\pm$6     &      292$\pm$6    &      292$\pm$6     &      292$\pm$6    &      292$\pm$6    &	   292$\pm$6    \\
6583.41       & {\fnii}  	& 1F   	       &   107.12$\pm$2.14&      113$\pm$2     &       122$\pm$2     &      123$\pm$2     &      124$\pm$2    &      130$\pm$3     &      135$\pm$3    &      148$\pm$3    &	   159$\pm$3    \\
6678.15       & {\hei}  	& 46  	       &     1.11$\pm$0.06&	1.13$\pm$0.06  &      0.89$\pm$0.04  &     0.20$\pm$0.01  &	    ---     & 	---      &	     ---     &     0.03$\pm$0.01 &	  0.26$\pm$0.01 \\
6716.47       & {\fsii}  	& 2F   	       &    18.79$\pm$0.94&     20.2$\pm$1.0   &      22.2$\pm$1.1   &     25.8$\pm$1.3   &     29.4$\pm$1.5  &     36.9$\pm$1.8   &	 42.3$\pm$2.1  &     49.1$\pm$2.5  &	  58.8$\pm$1.2  \\
6730.85       & {\fsii}  	& 2F   	       &    20.02$\pm$1.00&     21.4$\pm$1.1   &      23.0$\pm$1.2   &     26.6$\pm$1.3   &     30.4$\pm$1.5  &     38.7$\pm$1.9   &	 42.8$\pm$2.1  &     53.3$\pm$1.1  &	  60.8$\pm$1.2  \\
7065.28       & {\hei}  	& 10   	       &     0.92$\pm$0.05&	0.85$\pm$0.04  &      0.65$\pm$0.03  &     0.15$\pm$0.01  &	    ---     & 	---      &	 0.06$\pm$0.01 &     0.09$\pm$0.01 &	  0.34$\pm$0.02 \\
7135.78       & {\fariii}  	& 1F   	       &     2.90$\pm$0.14&	2.66$\pm$0.13  &      1.70$\pm$0.08  &     0.31$\pm$0.02  &	0.03$\pm$0.01 &     0.03$\pm$0.01  &	 0.19$\pm$0.01 &     0.24$\pm$0.01 &	  1.16$\pm$0.06 \\
7319.19       & {\foii}  	& 2F   	       &     1.74$\pm$0.09&	1.67$\pm$0.08  &      1.67$\pm$0.08  &     1.51$\pm$0.08  &	1.52$\pm$0.08 &     1.66$\pm$0.08  &	 1.80$\pm$0.09 &     2.53$\pm$0.13 &	  2.67$\pm$0.13 \\
7330.20       & {\foii}  	& 2F   	       &     1.46$\pm$0.07&	1.41$\pm$0.07  &      1.34$\pm$0.07  &     1.24$\pm$0.06  &	1.26$\pm$0.06 &     1.39$\pm$0.07  &	 1.50$\pm$0.08 &     2.07$\pm$0.10 &	  2.20$\pm$0.11 \\
\noalign{\vskip3pt} \noalign{\hrule} \noalign{\vskip3pt}
\multicolumn{3}{r}{$c$(H$\beta$)} 	& 0.60$\pm$0.09 & 0.49$\pm$0.08 & 0.51$\pm$0.05	& 0.75$\pm$0.06 & 0.76$\pm$0.06	& 0.79$\pm$0.07	& 0.77$\pm$0.06	& 0.73$\pm$0.06	& 0.99$\pm$0.07  \\
\multicolumn{3}{r}{$F$(H$\beta$)} & 1.508$\times$10$^{-12}$ 	& 1.443$\times$10$^{-12}$  	& 1.185$\times$10$^{-12}$   & 1.861$\times$10$^{-12}$  	&  1.770$\times$10$^{-12}$  & 1.765$\times$10$^{-12}$   &  1.271$\times$10$^{-12}$  &  1.150$\times$10$^{-12}$ &  7.131$\times$10$^{-13}$ \\
\noalign{\vskip3pt}
\multicolumn{3}{r}{\Ne(\foii)}& 560$\pm$50 & 520$\pm$50 	& 500$\pm$40	& 440$\pm$40	& 450$\pm$40 	& 450$\pm$40 	& 420$\pm$40	& 510$\pm$50	& 460$\pm$40   \\
\multicolumn{3}{r}{\Ne(\fsii)}& 650$\pm$170 	& 630$\pm$170 	& 570$\pm$160	& 560$\pm$150	& 560$\pm$150 	& 600$\pm$160 	& 520$\pm$150	& 690$\pm$140	& 580$\pm$60   \\
\multicolumn{3}{r}{\Te(\foii)}& 7850$\pm$160	& 7600$\pm$150 	& 7360$\pm$140	& 7260$\pm$130 	& 7270$\pm$140 	& 7420$\pm$140 	& 7440$\pm$140	& 8070$\pm$180	& 7810$\pm$180 \\
\multicolumn{3}{r}{\Te(\foiii)$^{(3)}$} & 7360$\pm$1700 & 7000$\pm$1650	& 6660$\pm$1600	& --- 	& --- 	& --- 	& --- 	& --- 	& --- 	\\
\noalign{\vskip3pt}
\multicolumn{3}{r}{He$^+$/H$^+$} 	&10.44$\pm$0.02 &10.45$\pm$0.03 &10.34$\pm$0.03	& --- 	& --- 	& --- 	& --- 	& --- 	& --- 	\\
\multicolumn{3}{r}{O$^+$/H$^+$} 	& 8.34$\pm$0.05 & 8.45$\pm$0.05 & 8.55$\pm$0.05	& 8.58$\pm$0.05 & 8.58$\pm$0.05 & 8.54$\pm$0.05 & 8.57$\pm$0.05 & 8.42$\pm$0.05	& 8.56$\pm$0.06 \\
\multicolumn{3}{r}{O$^{2+}$/H$^+$} 	& 6.97$\pm$0.20 & 6.64$\pm$0.22 & 5.83$\pm$0.26	& --- 	& --- 	& --- 	& --- 	& --- 	& --- 	\\
\multicolumn{3}{r}{N$^+$/H$^+$} 	& 7.63$\pm$0.04 & 7.70$\pm$0.04 & 7.78$\pm$0.04 & 7.80$\pm$0.04 & 7.80$\pm$0.04 & 7.80$\pm$0.04 & 7.81$\pm$0.04 & 7.73$\pm$0.04	& 7.81$\pm$0.04	\\
\multicolumn{3}{r}{S$^+$/H$^+$}	 	& 6.28$\pm$0.04 & 6.35$\pm$0.04 & 6.43$\pm$0.04 & 6.51$\pm$0.04 & 6.56$\pm$0.04 & 6.64$\pm$0.04 & 6.68$\pm$0.04 & 6.66$\pm$0.04	& 6.77$\pm$0.04 \\
\multicolumn{3}{r}{S$^{2+}$/H$^+$} 	& 6.78$\pm$0.04 & 6.80$\pm$0.04 & 6.87$\pm$0.04	& 6.83$\pm$0.04	& 6.73$\pm$0.04 & 6.64$\pm$0.04	& 6.66$\pm$0.04	& 6.39$\pm$0.05	& 6.44$\pm$0.05 \\
\multicolumn{3}{r}{Ar$^{2+}$/H$^+$} 	& 5.79$\pm$0.16 & 5.82$\pm$0.17	& 5.69$\pm$0.17	& --- 	& --- 	& --- 	& --- 	& --- 	& --- 	\\
\noalign{\vskip3pt}
\multicolumn{3}{r}{O/H} 		& 8.36$\pm$0.05 & 8.45$\pm$0.05 & 8.55$\pm$0.05 & 8.58$\pm$0.05 & 8.58$\pm$0.05 & 8.54$\pm$0.05 & 8.57$\pm$0.05	& 8.42$\pm$0.05	& 8.56$\pm$0.06	\\
\multicolumn{3}{r}{N/H} 		& 7.65$\pm$0.08 & 7.71$\pm$0.08 & 7.78$\pm$0.08 & 7.80$\pm$0.04 & 7.80$\pm$0.04 & 7.80$\pm$0.04 & 7.81$\pm$0.04	& 7.73$\pm$0.04	& 7.81$\pm$0.04	\\
\multicolumn{3}{r}{S/H}	 		& 6.90$\pm$0.04 & 6.93$\pm$0.03 & 7.01$\pm$0.03 & 7.00$\pm$0.03 & 6.96$\pm$0.03 & 6.94$\pm$0.03 & 6.97$\pm$0.03	& 6.85$\pm$0.03	& 6.94$\pm$0.03	\\
\noalign{\vskip3pt} \noalign{\hrule} \noalign{\vskip3pt}
\end{tabular}
\begin{description}
\scriptsize
\item[] $^{(1)}$ The errors in the line fluxes refer only to uncertainties in the line measurements (see text). $F$(H$\beta$) in erg\,cm$^{-2}$s$^{-1}$; 
\Ne\ in cm$^{-3}$; \Te\ in K; ionic abundances in log\,(X$^{+i}$/H$^+$)\,+\,12.
\item[] $^{(2)}$ Blended with {\hei} $\lambda$3833.57 line.
\item[] $^{(3)}$ \Te(\foiii) obtained using the empirical relation between \Te(\foii) and \Te(\foiii) obtained from the data by \citet{Gar07}
(see text).
\end{description}

\end{table*}
%

\subsection{Extinction correction}
\label{section43}
%
We assume the extinction law derived by \cite{Bla07} for
the Orion nebula. Both \hii\ regions have a deficiency in small particles that produces the relatively 
grey extinction observed in Orion \citep{Mag86, Bal91}. Furthermore, 
\citet{Rod99, Rod02} did not find large extinction variations in her optical spectroscopic studies 
of the M\,42 and M\,43 \hii\ regions.

The reddening coefficient, $c$({\hb}) was obtained by fitting the observed 
H$\delta$/{\hb} and H$\gamma$/{\hb} line intensity ratios --\,the three lines lie in
the same spectral range\,-- to the theoretical ones computed by \cite{Sto95} for 
{\te}\,=\,7500 K and {\Ne}\,=\,500 cm$^{-3}$. As pointed out by 
\cite{Mes09a, Mes09b}, the use of a \cite{Bla07} law instead of the classical one 
by \cite{Cos70} produces slightly higher 
$c$({\hb}) values and also slightly different dereddened line fluxes depending on 
the spectral range.

The final set of line intensities corrected for both the extended emission component
and the extinction are indicated in Table~\ref{t2}.

\subsection{Uncertainties}
\label{section44}

Several sources of uncertainties must be taken into account to obtain
the errors associated with the line intensity ratios. We estimated that 
the uncertainty\footnote{Although uncertainties in the line flux measurements depend 
on the line flux instead of on $F$($\lambda$)/$F$({\hb}), the similarity between 
the {\hb} flux measured in each aperture (see Table~\ref{t2}) makes this approach 
reasonable.} in the line intensity measurement due to the signal-to-noise ratio
of the spectra and the placement of the local continuum is typically
\begin{itemize}
\item [$\sim$]\,2\% for $F$($\lambda$)/$F$({\hb})$\geq$0.5, 
\item [$\sim$]\,5\% for 0.1$\leq$ $F$($\lambda$)/$F$({\hb}) $\leq$0.5, 
\item [$\sim$]\,10\% for 0.05$\leq$ $F$($\lambda$)/$F$({\hb}) $\leq$0.1, 
\item [$\sim$]\,20\% for 0.01$\leq$ $F$($\lambda$)/$F$({\hb})$\leq$0.05, 
\item [$\sim$]\,30\% for 0.005$\leq$ $F$($\lambda$)/$F$({\hb})$\leq$0.01, and 
\item [$\sim$]\,40\% for 0.001$\leq$ $F$($\lambda$)/$F$({\hb})$\leq$0.005.  
\end{itemize}
We did not consider any lines of weaker intensity than 0.001 $\times$ $F$({\hb}).
We note that then uncertainties indicated in Table \ref{t2} refer inly to this type 
of errors.
 
By comparing the resulting flux-calibrated spectra of one of our standard stars 
with the corresponding tabulated flux, we could estimate that the line ratio uncertainties 
associated with the flux calibration is 
$\sim$\,3\% when the wavelengths are separated by 500\,--\,1500 \AA\  and $\sim$\,5\% 
if they are separated by more than that. Where 
the corresponding lines are separated by less than 500 \AA, the uncertainty 
in the line ratio due to uncertainties in the flux calibration is negligible.
 
The uncertainty associated with extinction correction was computed by error propagation.
The contribution of this uncertainty in the total error is again negligible when
line ratios of close-by lines are considered 
(e.g. [\ion{S}{ii}]\,$\lambda$6716/[\ion{S}{ii}]\,$\lambda$6730).
The final errors in the line intensity ratios used to derive the physical properties
of the nebula were computed by quadratically adding these three sources of uncertainty. 

\subsection{Physical conditions}
\label{section45}
%
The electron temperature (\te) and density (\Ne) of the ionized gas were
derived from classical CEL ratios, using the \iraf\ task
\emph{temden} of the package \emph{nebular} \citep{Sha95} with updated 
atomic data for several ions \cite[see][]{Gar09}. We computed {\Ne} from
the {\fsii} $\lambda$6717/$\lambda$6731 and {\foii} $\lambda$3729/$\lambda$3726
line ratios and {\te} from the nebular to auroral
{\foii} $\lambda\lambda$(7320+30)/$\lambda\lambda$(3726+29) line ratio\footnote{In 
the case of M\,43, \te({\foii}) is free of any contamination to 
{\foii} $\lambda$7325 by recombination because of the low \ion{O}{$^{2+}$} abundance.}.

We could not directly determine {\te}({\foiii}) because
the {\foiii} $\lambda$4363 auroral line was not detectedbecause of the low ionization degree
of the nebula. As an alternative, we used a empirical relation between {\te}({\foiii}) and 
{\te}({\foii}) obtained from the nebular information published by \citet{Gar07} for a sample of 
Galactic {\hii} regions\footnote{The data for NGC3603, a giant Galactic {\hii} region, was discarded
to obtain this relation.} 
\begin{equation}
\label{te_rel}
T_e({\rm [O\,III]})=[T_e({\rm [O\,II]})-(2640\pm1270)]/(0.7105\pm0.1524).
\end{equation}
This fit is almost identical to that obtained by \citet{Pil06} and \citet{Gar92} 
for integrated spectra of giant extragalactic {\hii} regions and 
photoionization models of giant extragalactic {\hii} regions, respectively. 
We only computed {\te}({\foiii}) for the three innermost apertures. For more
distant apertures, the O$^{2+}$ emission (if any) is not associated to
M\,43 itself, but to the residual emission of the diffuse component (see Sects. \ref{section3}
and \ref{section41}).

The methodology used to determine the physical conditions was as follows:
we assumed a representative initial value of {\te} of 10000 K and computed the
electron densities. The value of {\Ne} was then used to compute 
{\te}({\foii}) from the observed line ratios, and {\te}({\foiii}) from equation 2. 
We iterated until convergence to obtain the final values of {\Ne}  and {\te}. 
Uncertainties were computed by error propagation. The final {\Ne}({\foii}), 
{\Ne}({\fsii}), {\te}({\foii}), {\te}({\foiii}) estimate, along with their uncertainties
are indicated in Table~\ref{t2}.

In general, densities derived from the {\foii} line ratio are about 100 cm$^{-3}$
lower than those derived from {\fsii} lines, but consistent within the
errors. 

%
\subsection{Chemical abundances}
\label{section46}
%
To derive He$^+$/H$^+$, we used two observed lines of {\hei} at $\lambda\lambda$4471 and 6678.
Case B emissivities were taken from the collision less (low-density limit) calculations by
\citet{Bau05} using an on line available code\footnote{Available at \tt{http://www.pauKy.edu/$\sim$rporter/j-resolved}}.
The collisional-to-recombination contribution was estimated from \citet{Kin95}, using the interpolation
formula provided by \citet{Por07}. The effective recombination coefficients for H$^+$ were taken
from \citet{Sto95}

Ionic abundances of N$^+$, O$^+$, O$^{2+}$, S$^+$, S$^{2+}$ and Ar$^{2+}$ were
derived from CELs, using the \iraf\ task $ionic$ of the package $nebular$. We
assumed a two-zone scheme, in which we adopted {\te}({\foii}) for the low
ionization potential ions (N$^+$, O$^+$, S$^+$ and S$^{2+}$) and the value of {\te}({\foiii}) derived
from Eq.~\ref{te_rel} for the high ionization potential ions (O$^{2+}$ and Ar$^{2+}$).
In both cases, we assumed {\Ne}({\foii}) instead of {\Ne}({\fsii}) because the first indicator 
is more representative of the whole nebula, while {\Ne}({\fsii}) provides insight mostly 
into the density of the material close to the ionization edge of the nebula. 
For O$^{2+}$, Ar$^{2+}$, and He$^+$, we only have available lines in the three innermost apertures. 
Owing to the low \Teff\ of the ionizing star, these relatively high ionization species 
are only present in regions very close to the star.

We then derived the total abundances of O, N, and S for each aperture\footnote{Owing to the 
low ionization degree of M\,43, the ionization correction factors for He$^+$ and Ar$^{2+}$ are 
large and very uncertain. We did not 
compute the total abundances for these two elements.}.
As a consequence of the low ionization degree of the nebula, M\,43 has a very small amount of O$^{2+}$ 
in comparison with the dominant species, O$^{+}$, except in the innermost zones where a non-negligible fraction of 
O$^{2+}$ is present. We note that for the intermediate apertures (apertures 4 to 6), 
the O abundance is given by O$^+$/H$^+$. In the case of S, the abundance of this element is given by the sum of
S$^+$ and S$^{2+}$, because S$^{+3}$ is not expected to be present in this nebula. Similarly to 
O$^{2+}$, N$^{2+}$ is only expected to be present in the innermost parts of the nebula. For the corresponding 
apertures (1\,--\,3), we applied the classical ionization correction factor (ICF) scheme of N (i.e. N$^+$/O$^+$=N/O) to correct for the
N$^{2+}$ abundance (note that this correction is very small). As for O, for the intermediate 
apertures the N abundance is directly obtained from the N$^+$ abundance.

The ionic abundances of He$^{+}$ O$^{+}$, O$^{2+}$, N$^{+}$, S$^{+}$, S$^{2+}$, and Ar$^{2+}$, along
with the total abundances of O, N, and S are shown in Table~\ref{t2}. We stress again that 
in this nebula the total abundances of O, N, and S can be obtained directly from observable ions 
(no ICFs are needed) from the intermediate apertures. 

\subsection{Comparison of O, N, and S abundances derived in M\,43 and the Orion nebula}
\label{section47}
%
Given that the ionized gas from both nebulae, M42 and M43, comes from the 
same molecular cloud, it is reasonable to assume that both have very similar 
total abundances. In this section, we compare the O, S, and N abundances obtained 
from our study of M\,43 with those determined by \citet[][GRE07, an update of Esteban et al. 2004]{Gar07} 
and \citet[][SDS10]{Sim10b} 
from the analysis of the nebular spectrum of the Orion nebula. Table \ref{t3}
summarizes the final abundances indicated in the three studies. We note that
GRE07, and SDS10 present analyses of the same spectroscopic
dataset, but use different atomic data and ICFs. Our study of M\,43 uses the same
atomic data as that considered by GRE07.

\subsubsection{Oxygen}
\label{section471}

For this element, where the total abundance is obtained directly from observable
ions (\ion{O}{$^+$} and \ion{O}{$^+$}+\ion{O}{$^{2+}$} in the case of M\,43
and M\,42, respectively), good agreement is found within the errors. 

\subsubsection{Nitrogen}
\label{section472}

%
%
\begin{table}[t!]
\begin{center}
\caption{\footnotesize Summary of O, N, and S abundances (from CEL) resulting from
the spectroscopic analysis of M\,43 (this work) and the Orion nebula (GRE07, SDS10). 
For M\,43, we assumed the mean value of abundances derived from apertures 4\,--\,6 (the
uncertainty given by the values indicated in Table \ref{t2} for each aperture).  
\label{t3}
}
\begin{tabular}{cccc}    
\hline
\hline
\noalign{\vskip2pt}
Element & M\,43 & M\,42 (GRE07) & M\,42 (SDS10) \\
\noalign{\vskip1pt}
\hline
O & 8.57$\pm$0.05  & 8.54$\pm$0.03   &  8.52$\pm$0.01 \\
S & 6.97$\pm$0.03  & 7.04$\pm$0.04   &  6.87$\pm$0.04 \\
N & 7.80$\pm$0.04  & 7.73$\pm$0.09   &  7.90$\pm$0.09 \\
\hline
\hline
\end{tabular}
\\
\end{center}
\end{table}

For this element, GRE07 and SDS10 evaluate different values.
This discrepancy is caused by the assumed ICF(N$^+$). While GRE07
obtained this ICF from the widely used empirical relation N$^+$/O$^+$=N/O, 
SDS10 determined the ICF from a model fitted to the data of M\,42
\citep[including a detailed description of the ionizing SED of $\Theta^1$\,Ori\,C
predicted by the stellar atmosphere code WM-basic, ][]{Pau01}.
The GRE07 and SDS10 abundances differ by 0.17 dex. Interestingly, our determination for
M\,43, in which no ICF is needed produces an intermediate value, and is in 
perfect agreement with the B-type stars abundance derived by \citet{Nie11}. 
This introduces a new important constraint on the nebular CEL/RL abundance 
conundrum \citep{Sim10b}.

\subsubsection{Sulphur}
\label{section472}

GRE07 and SDS10 derive different
values (0.17 dex difference) of the total S abundance in M\,42. In
this case, the cause of the difference is not, however, the assumed ICF, but the use
of different sets of atomic data. Both computations considered the same 
data for the collision strength \citep{Kee96, Tay99}, but different transition 
probabilities: while GRE07 based their calculation in data from \cite{Men82a, Men82b}
for [\ion{S}{ii}] and [\ion{S}{iii}], respectively, SDS10 used \cite{Fro04a} and \cite{Fro04b} 
values. For M\,43, we obtained 12\,+\,log(S/H)=6.97$\pm$0.04 using the 
same set of atomic data as GRE07. This value is slightly lower, but in agreement within 
the errors, than the determination by GRE07, and somewhat
higher than the value provided by SDS10. The influence
of the atomic data considered for this element is hence relevant.

\subsection{Comparison with previous studies of M\,43}
\label{section48}

The only nebular abundance studies found in the literature 
for M\,43 are those of \citet{Rod99, Rod02}, who 
analyzed long-slit optical spectroscopic data for five different slit 
positions along the nebula, deriving the physical conditions (\Te\ 
and \Ne), and O, S, Cl, N, Ar, He, C, and Fe abundances of the nebular 
gas. \citet{Ode10} analyzed a set of optical spectrophotometric 
observations of M\,43 as part of a more general study of the EON, and derived the 
physical conditions of the nebula. We note that none of these observations were corrected 
for the diffuse emission component as in our study.

For consistency, we reanalyzed the data of \citet{Rod99} using the same atomic data as 
in our work. We obtained {\te}({\fnii}) between 7800 K and 8000 K for slit positions similar 
to ours. \citet{Ode10} obtained very similar results ({\te}({\fnii}) between 7780 K and 7940 K) 
from several aperture extractions of a slit that crosses the nebula in the east-west direction. 
Both authors derived \Ne\ ({\fsii}) $\sim$500-600 cm$^{-3}$, which is very similar to the 
value derived in this work. 

We compared our abundances with those resulting from the reanalysis\footnote{The observations by 
\citet{Ode10} do not include {\foii}$\lambda\lambda$3727, 7325 lines, 
hence the determination of the oxygen abundance is not possible.} of the observational
dataset by \citet{Rod99}. The derived O abundances are 0.14\,--\,0.23 dex lower 
than those resulting from the analysis of our spectra. As we indicate in Sect. \ref{section49},
this effect is produced by the contamination of the M\,43 emission by the diffuse component. 

Finally, we compared our estimated {\te}({\foii}) with previous determinations
from radio continuum observations. \cite{Sub92} derived
\Te=\solu{9000}{1700} K by combining 330 MHz and 10.7 GHz continuum observations and a
model of an isothermal \hii\ region; \cite{Mil68} estimated a value of \Te$\sim$8000 K from 408 MHz
continuum observations; finally, \cite{Thu78} derived a much lower \Te$\sim$6700 K
from the HI 91$\alpha$ radio recombination line to continuum ratio. In general, the different estimates, 
except for Thum et al., agree within the uncertainties. 

%
%

\begin{figure}[!t]
\centering
\includegraphics[width=8.5cm,angle=0]
{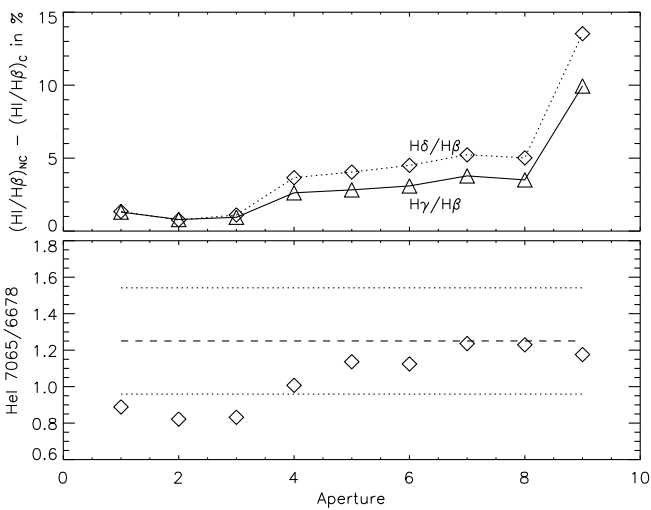}
\caption{Upper panel: Difference between the non-extinction-corrected H$\gamma$/{\hb} and H$\delta$/{\hb} ratios from
the C and NC spectra (see text for explanation).
Lower panel: {\hei} 7065/6678 line ratio from the NC spectra (9 apertures). The mean and standard
deviation values corresponding to the bright Huygens region \citep[obtained from data by][]{Ode10}
are also indicated as a dashed and dotted lines, respectively.}
\label{f8}
\end{figure}

%
%
\begin{figure*}[!t]
\centering
\includegraphics[angle=90, width=18cm]
{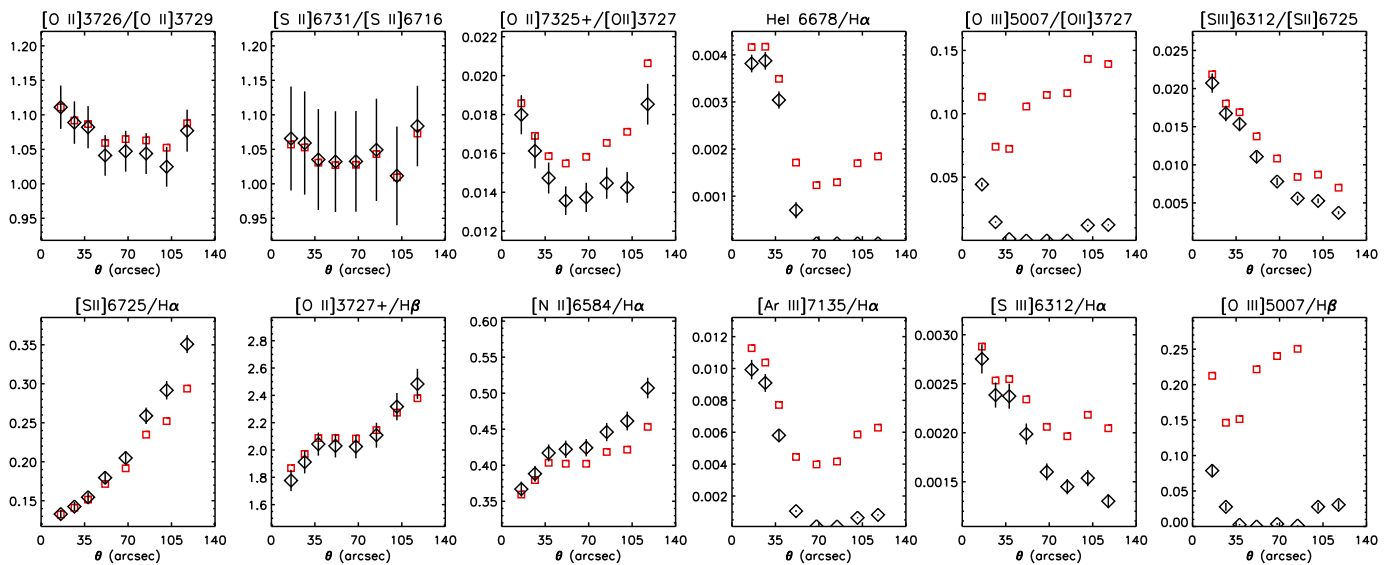}
\caption{Observed line ratios obtained from the corrected (black diamonds) and 
non corrected (red squares) spectra from extended emission component.}
\label{f9}
\end{figure*}

\subsection{Scattered nebular light from the EON affecting the spectrum of M\,43?}
\label{section49}
%

\citet{Ode09} and \citet{Ode10} demonstrated that scattered light from
the bright Huygens Region of M\,42 can affect the physical conditions derived from emission
lines in the EON. \citet{Ode10} performed a detailed spectroscopic study of several zones
of the EON, including the M\,43 region. These authors could not find conclusive
evidence of signatures of scattered light from the Huygens region affecting the nebular 
emission from M\,43 (except for the case of an aperture located in the M\,43 
dark lane). Our study (based on more detailed observational dataset) allowed us 
to observationally prove that there is an extended diffuse emission not produced by 
ionization of the surrounding material by NU Ori affecting the whole nebula.

To investigate whether this extended emission relates to scattered light from the 
Huygens region, we compare in Fig.~\ref{f8} (upper panel) the 
non-extinction-corrected H$\gamma$/{\hb} and H$\delta$/{\hb} line ratios with (C) and without (NC) correction 
for diffuse emission for apertures from A1 to A9. The H$\delta$/{\hb} 
line ratio is more affected than 
H$\gamma$/{\hb}. This is consistent with the diffuse extended emission being produced by
dust scattered light, a well-know property of this type of emission is that blue
lines are more affected than red lines. An additional argument supporting this hypothesis
is presented in the bottom panel of Fig.~\ref{f8}. The value of the 
{\hei} 7065/6678 ratio, obtained from the NC spectra of M\,43 outside the \ion{He}{$^+$} 
Str\"omgen sphere associated with NU Ori
agrees with the associated average value in the Huygens region taken from the data of 
\citet{Ode10}\footnote{We considered extractions corresponding to slits 2 to 15 that are 
labeled as ''inner''.}.

We also investigated the effect that scattered light contaminating the spectrum of
M\,43 has on line ratios used to determine the physical conditions and chemical
abundances (as well as constraints of the photoionization models presented in
Paper~II). Fig.~\ref{f9} shows several line ratios obtained from the
C and NC spectra. Although the effect is negligible for the density constraints 
and the \ion{S}{$^+$} and \ion{O}{$^+$} abundance indicators, the influence of
scattered light on the ionization structure constraints and the \ion{He}{$^+$},
\ion{N}{$^+$}, \ion{Ar}{$^2+$}, \ion{S}{$^2+$}, and \ion{O}{$^{2+}$} abundance indicators
is not negligible (especially in the outer parts of the nebula). In addition, as already pointed out by 
\citet{Ode10}, our comparison identifies an important effect on \Te([\ion{O}{ii}]).
If the scattered light contribution is not removed, the spectroscopic analysis 
overestimates the \Te\ and hence underestimates of the abundances. 

To quantitatively illustrate this effect on the computed physical conditions and 
abundances, we present in Table \ref{t4} the results of two empirical analyses of 
the C and NC spectra from apertures 4 and 5. We also compare the result of the 
analysis of the NC integrated spectrum (extracting the global long\,slit spectra). 
As pointed out before, \Ne\ is basically similar but the NC spectra result in a 
higher \te\ and, consequently lower abundances. In particular, ionic abundances can 
be affected by $\sim$\,0.05\,--\,0.15 dex.

An adequate correction of the scattered light contribution in the spectra of M\,43 is hence
important for a reliable empirical analysis of the nebular spectra of
this \hii\ region.


\begin{table}[t!]
\begin{center}
\caption{\footnotesize Summary of physical conditions, ionic and total abundances derived 
for apertures 4 and 5 with and without correcting for 
extended emission component and for the longslit extraction in M\,43.$^{(1)}$}
\label{t4}
\begin{tabular}{l l c c c c c}    
\hline
& & \multicolumn{2}{c}{Uncorrected} & \multicolumn{2}{c}{Corrected} &  Uncorrected \\
\cline{3-6}
& & A4 &  A5 &  A4 &  A5 & Total Slit \\
\cline{3-7}
\Ne (cm$^{-3}$)		& {\foii}	& 475 	& 485	& 440	& 450	& 510	\\
                   	& {\fsii}	& 560 	& 565	& 560	& 560	& 600	\\
\hline
\Te (K)			& {\foii}	& 7650 	& 7700	& 7260	& 7270	&  7900 \\
			& {\foiii}	& --- 	& ---	& ---	& ---	&  7450 \\
\hline
$\epsilon$(X$^{+i}$)	& O$^+$		& 8.45 	& 8.44	& 8.58	& 8.58	&  8.38 \\
			& O$^{++}$	& --- 	& ---	& ---	& ---	&  7.45 \\
			& N$^{+}$	& 7.71 	& 7.70	& 7.80	& 7.80	&  7.66 \\
			& S$^{+}$	& 6.42 	& 6.46	& 6.51	& 6.56	&  6.42 \\
			& S$^{++}$	& 6.77 	& 6.70	& 6.83	& 6.73	&  6.69 \\
			& Ar$^{++}$	& --- 	& ---	& ---	& ---	&  5.64 \\
\hline
$\epsilon$(X)   	& O		& 8.50 	& 8.49	& 8.58	& 8.58	&  8.43 \\
			& N		& 7.75 	& 7.75	& 7.80	& 7.80	&  7.71 \\
			& S		& 6.93 	& 6.90	& 7.00	& 6.96	&  6.88 \\
\hline
\end{tabular}
\begin{description}
\scriptsize
\item[] $^{(1)}$ \Te, ionic, and total abundance values are based on the assumption
 \Ne=\Ne({\foii}) as discussed in Sect. \ref{section46}.
\end{description}
\end{center}
\end{table}

%
\section{Summary and conclusions}
\label{section7}
%
M\,43 is a close-by, bright \hii\ region of simple
geometry, ionized by a single star. These characteristics makes M\,43 an
ideal object for investigating several topics of interest in the field
of \hii\ regions and massive stars.

In a series of two papers, we present a combined, comprehensive study of 
the nebula and its ionizing star using as many observational constraints 
as possible. In this first part of the study, we have introduced the observational 
dataset, obtained the stellar parameters of HD\,37061, obtained useful information 
from the nebular images, and analyzed the nebular spectra
extracted from apertures located at various distances from the central star.
All this information is then used in Paper II to construct a customized 
photoionization model of the nebula. \\

In this paper, we have found observational evidence of a diffuse and extended emission 
in the region where M\,43 is located that is not associated to material ionized 
by HD\,37061. Our nebular observations have allowed us to ascertain the most likely 
origin of this emission \citep[e.g.][and references therein]{Ode09, Ode10}, namely 
that light emitted in the Huygens region (the central, brightest part of the Orion nebula) 
is scattered by dust. We have also shown the importance of an adequate correction
of this scattered light from the imagery and spectroscopic observations of M\,43
to a proper determination of the total nebular \ha\ luminosity, the nebular physical
conditions, and chemical abundances. In particular, we illustrated that an overestimate of 
\Te\ by $\sim$\,400\,--\,500 K, hence an underestimate of abundances by 
0.05\,--\,0.10 dex, result from the empirical analysis of our spectroscopic dataset
when the spectra are not corrected for the scattered light contribution. In
addition, the derived total nebular \ha\ luminosity may be overestimated by a factor
$\sim$\,1.5.

The quantitative analysis of the optical spectrum of HD\,37061 with the 
stellar atmosphere code FASTWIND lead to \Teff\,=\,31000$\pm$500 K, \grav\,=\,4.2$\pm$0.1, and 
log\,Q(H$^0$)\,=\,47.2\,$\pm$\,0.2 (assuming a distance of 400\,pc).
 
The analysis of the \ha\ and \hb\ images indicate a non-constant extinction distribution
within the nebula that is well correlated with the dust features indicated by 
\cite{Smi87}. Once the \ha\ image is corrected for extinction and diffuse emission, 
a total nebular \ha\ luminosity of (3.0$\pm$1.1) x 10$^{35}$ erg\,s$^{-1}$ is obtained. 
This value is compatible with the ionizing flux of the star, implying that the nebula is 
(mostly) ionization bounded.

We extracted nine apertures from a long-slit located to the west of HD\,37061 (east-west direction) 
to obtain the spatial distribution of the nebular physical conditions (temperature and density) 
and ionic abundances (\ion{He}{$^+$}, \ion{O}{$^+$}, \ion{O}{$^{2+}$}, \ion{N}{$^+$}, 
\ion{S}{$^+$}, \ion{S}{$^{2+}$}, \ion{Ar}{$^{2+}$}). Since it is important to correct
these spectra for the contribution of scattered light from the Huygens region, we also consider 
two apertures outside the nebula that are used for this aim. 

For sulfur, oxygen, and nitrogen, we have been able to determied total abundances directly 
from observable ions (no ICFs are needed). The derived abundances, 8.57\,$\pm$\,0.05, 
6.97\,$\pm$\,0.03, and 7.80\,$\pm$\,0.04, respectively, are compared with previous 
determinations in the Orion nebula. Although an overall agreement is found, our study illustrates 
the importance of the atomic data and, specially in the case of M\,42, the considered ICFs.

\begin{acknowledgements}
%
SSD acknowledges the funds by the Spanish Ministerio de Educaci\'on y Ciencia 
under the MEC/Fullbright postdoctoral fellowship program.
SSD, JGR, CE, and ARLS acknowledge finantial support by the Spanish MICINN under 
projects AYA2008-06166-C03-01 and AYA2007-63030. SSD also acknowledges financial 
support from UNAM/DGAPA PAPIIT IN 112708 (IP: M. Pe\~na) and CONACyT J49737 
(IP: C. Morisset) and the members of the Instituto de Astronom{\'{\i}}a, 
UNAM for their warm hospitality. JGR acknowledges the support of an UNAM postdoctoral grant. 
Part of this work was done while CM in sabbatical at IAC founded by CONACyT and PASPA-UNAM grants.
This work has also been partially funded by the Spanish MICINN under the 
Consolider-Ingenio 2010 Program grant CSD2006-00070: First Science with the GTC.
%
\end{acknowledgements}
%

%
%


%
\end{document}